\documentclass[format=acmsmall, review=false, screen=true]{acmart}

\usepackage{booktabs} 
\usepackage{enumerate}
\usepackage[ruled]{algorithm2e} 

\SetAlFnt{\small}
\SetAlCapFnt{\small}
\SetAlCapNameFnt{\small}
\SetAlCapHSkip{0pt}
\IncMargin{-\parindent}

\begin{document}
\title[Structured Query Engine]{Building a Structured Query Engine}  
\author{Amanpreet Singh, Karthik Venkatesan and Simranjyot Singh Gill}
\orcid{}
\affiliation{%
  \institution{New York University}
  \streetaddress{251 Mercer St}
  \city{New York}
  \state{New York}
  \postcode{10012}
  \country{USA}}

\begin{abstract}
Finding patterns in data and being able to retrieve information from those patterns is an important task in Information retrieval. Complex search requirements which are not fulfilled by simple string matching and require exploring certain patterns in data demand a better query engine that can support searching via structured queries. In this article, we built a structured query engine which supports searching data through structured queries on the lines of ElasticSearch. We will show how we achieved real time indexing and retrieving of data through a RESTful API and how complex queries can be created and processed using efficient data structures we created for storing the data in structured way. Finally, we will conclude with an example of movie recommendation system built on top of this query engine.
\end{abstract}

\maketitle

\renewcommand{\shortauthors}{G. Zhou et al.}

\section{Introduction}

Information retrieval (IR) deals with the representation, storage, organization of, and access to information items. The representation and organization of the information items should provide the user with easy access to the information in which he is interested. Normally, simple term based indexing can only provide matching with a specific term and return all the documents that contain that term. Usually, this doesn't meet the real needs of users, for e.g. a user might be looking for articles containing term "billionaire" in category of "sports". A normal match based search will return all the articles from all categories including "politics", "technology", "sports" and others that contain word "billionaire" which will not meet the needs of user. A better search engine that support searching in various categories and relatively more complex queries (union, intersection) on patterns is therefore required \cite{Brin:1998:ALH:297810.297827}. \\

	Elasticsearch\cite{Gormley:2015:EDG:2904394} is an open source a distributed, RESTful search engine which supports structured querying. Elasticsearch is built on top of Lucene which is a key value data store that uses inverted index \cite{Zobel:2006:IFT:1132956.1132959} as its main data structure. Elasticsearch provides a layer over lucene with which we can easily query complex patterns through building an AST of queries using its Query DSL \cite{ElasticQueryDSL}. We built a lucene like underlying indexer and Query DSL type retriever for structured query engine \footnote{Source code for the structured query engine is available at \href{http://github.com/apsdehal/structured-query-engine}{Github}, which can be used as a code reference to this article. We will be linking to specific code portions of project in this articles}, which will be referred to as "SQE" in this article. These were connected further through a RESTful API. We used ElasticSearch as a model to build SQE supporting its various important features. We will start by explaining the internal architecture of SQE, then dive into its various components and finally, we conclude with achievements and shortcomings of SQE.

\section{Structured Query Engine}
\subsection{Architecture}

\begin{center}
\includegraphics[width=10cm,height=10cm,keepaspectratio]{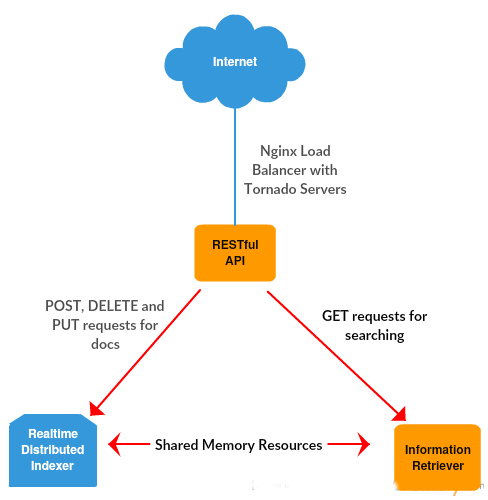}
\par Figure 1. Architecture of SQE
\end{center}

SQE's architecture has three main components: (1) RESTful API\cite{W3Rest} (2) Realtime Distributed Indexer (3) Information Retriever. Service is exposed to the world via a nginx load balancer which balances between several tornado\footnote{See \href{https://tornadoweb.org}{more} about Tornado web framework} servers which share the memory resources of indexers and retrievers between them. Servers are bind to localhost unless a person changes the bind address in the config. As with Elasticsearch, most of the parameters for SQE are configurable via service's config and particular index's settings and mappings. \\

	Architecture also contains a bootstrapper\footnote{\href{https://github.com/apsdehal/structured-query-engine/blob/master/app/helpers/utils/Bootstrapper.py}{Code} for bootstrapper} that bootstraps the whole service. It checks for the present configuration and indices and loads necessary datastores into memory to initialize indexers and retrievers for each index. It also provides a configuration module for all of the other modules to use which contains necessary information and objects. RESTful API takes in POST, PUT and DELETE requests for documents and passes them to Indexer which handles the indexing of the documents according to the mapping of the index. API also handles the GET requests for document id and search and passes them to indexer's and retriever's object respectively. URL scheme for the RESTful API is same as that of the Elasticsearch\cite{ElasticQueryDSL}. Inverted index and document store data structures are shared between a particular index's indexers and retrievers in order to avoid duplicate copies and keep all of them in synchronization.
    
\subsection{RESTful API}
RESTful API is the component of SQE which faces the internet. In our code base, we have implemented the API using Tornado Web Framework. API is divided into three main controllers\footnote{\href{https://github.com/apsdehal/structured-query-engine/tree/master/app/server/frontend/handlers}{Codebase} for controllers}: (1) Information controller (2) Index controller (3) Index query controller. Info controller provides the information about the current running service. Index controller handles the creation and deletion of indices. It handles creating the disk structure for an index, adhering to its settings as passed in the HTTP request body and saving its mappings and other necessary information. Index query controller handles the requests for searching, adding, updating or deleting documents into a particular index's types. This controller delegates the request for searching to information retriever and other requests to real-time distributed indexer.

\subsubsection{Mappings and Settings:} SQE supports most of the settings and mappings configuration of Elasticsearch. Number of shards are configurable for an index. Mapping\footnote{Mappings can be used to provide structure to data and how it is indexed. For more information on elasticsearch mapping see \href{https://www.elastic.co/guide/en/elasticsearch/reference/current/mapping.html}{documentation}} is the process of defining how a document, and the fields it contains, are stored and indexed. These mappings are flattened\footnote{\href{https://github.com/apsdehal/structured-query-engine/blob/master/app/indexer/Flattener.py}{Code} for flattener. For more information how the a mapping looks like when flattened see \href{https://www.elastic.co/guide/en/elasticsearch/reference/current/nested.html}{elasticsearch documentation}} so that the nested objects can be represented as single field\footnote{Index is parent in the hierarchy, which is followed by type object and then fields which have their own datatypes including "nested". For example, in movie recommendations, index is $movie\_recommendations$, object type is $movie$ and fields are $actor\_names$ and others. See \href{https://www.elastic.co/guide/en/elasticsearch/reference/current/mapping-types.html}{documentation} for field datatypes.} using "." for concatenation. A nested datatype field "commented" with "name" field will become "comment.name" and a separate inverted index will be generated for this field \cite{Lee:1996:ISS:226931.226950}. The same flattened field will be used as $field$ parameter while querying from information retriever. SQE supports various data-types from elasticsearch for its fields in mapping. These include:
\begin{enumerate}
\item Primitive data-type: This types include primitive data-types like integer, float and double. These are stored as it is without any analyzing.
\item Text data-type: These are analyzed into tokens before storing by specifying an analyzer property for a particular field. Default analyzer is "standard".
\item Keyword data-type: These are also fed directly to indexer without any kind of analyzing and are of type string.
\end{enumerate}

	SQE supports an $analyzer$ and $search\_analyzer$ properties for a field. $analyzer$ is the type of analyzer (tokenizer) to be used to generate tokens for inverted index while indexing. $search\_analyzer$ is the analyzer to be used to generate tokens of the query while calculating score of documents for a particular search query.
    
\subsection{Real-time Distributed Indexer}
The Indexer \cite{BottomUp} is a real-time, on-line component which is responsible for indexing the documents. It runs on multiple servers which are load balanced using Nginx. Each Indexer is further divided into multiple shards, the basic blocks of data, according to the document-partition scheme. This is required to parallelize the indexing and retrieval process for the SQE.
\subsubsection{Flatten Mapping}:
First, the mapping is flattened to remove nesting of fields. This makes it simpler to index the document and helps to retain the relationships between the fields within the documents.
\subsubsection{Analyzers:} Further, fields are analyzed to tokenize using the analyzer\footnote{See more analyzer \href{https://www.elastic.co/guide/en/elasticsearch/reference/current/analysis-analyzers.html}{here}} specified in "analyzer" property of the field. If "analyzer" is not present, then standard analyzer is used. The types of analyzer present for tokenizing in SQE are:
\begin{enumerate}
\item Standard: The standard analyzer divides text into terms on word boundaries using tokenize from nltk\_tokenize. It removes most punctuation, lowercases terms and english stopwords.
\item Whitespace: The whitespace analyzer divides text into terms whenever it encounters any whitespace character. It does not lowercase text.
\item Simple: The simple analyzer divides text into terms whenever it encounters a character which is not a letter. It lowercases all terms.
\item Ngram: Use "n\_gram" in analyzer field to use this analyzer. The ngram analyzer generates ngrams for a string's token attained through whitespace tokenizing. Minimum size of ngram is 3. This analyzer is good for implementing partial search with terms, but be careful for using it with search analyzer as you might not want to tokenize your search query into ngrams. 
\end{enumerate}
\subsubsection{Inverted Index}
The inverted index is the most important part of the SQE which allows for fast full text searches. The overall mapping is captured as a data structure of nested dictionaries. There is an index data structure for each $\_index$ separately. At the top level of each index data structure, the key is the $\_type$. At the second level, the mapping to the different shards takes places according to the hash of the $\_id$. The next level has the $fields$ which are indexed and the final level is the actual inverted index dictionary. It has the term as the key and gives a pair containing the document frequency and the postings dictionary. The postings dictionary provides a mapping from the documents the term is present to their term frequency. 
\subsubsection{Document Store}
The document store has all the documents which are present in the SQE. It is implemented as nested dictionaries with a similar structure to the index data structure until the shard number. The final level is a dictionary that takes the $\_id$ as the key and returns the corresponding document. The document store has a key called $num\_docs$ which gives the number of documents in the document store.
\subsubsection{Operations} The indexer supports the main operations of GET, PUT, POST and DELETE from HTTP requests in real-time.
\begin{enumerate}
\item \href{https://github.com/apsdehal/structured-query-engine/blob/master/app/indexer/Indexer.py\#L96}{Add}:
\newline
Each document is uniquely identified by its $\_id$. The user has the option to specify the $\_id$ parameter or the API generates one on its own. A top level entry is made in the index data structure if a new $\_type$ is encountered. The document is then flattened by the calling $Flatten Mapping$ and tokenized according to the $Analyzer$. This output is used to call the $generate$ function to add to the inverted index and the document store.Finally inverted index and document store are written to disc using a \href{https://github.com/apsdehal/structured-query-engine/blob/master/app/indexer/Indexer.py\#L208}{flush} operation.
\begin{center}
\includegraphics[width=13cm,height=10cm,keepaspectratio]{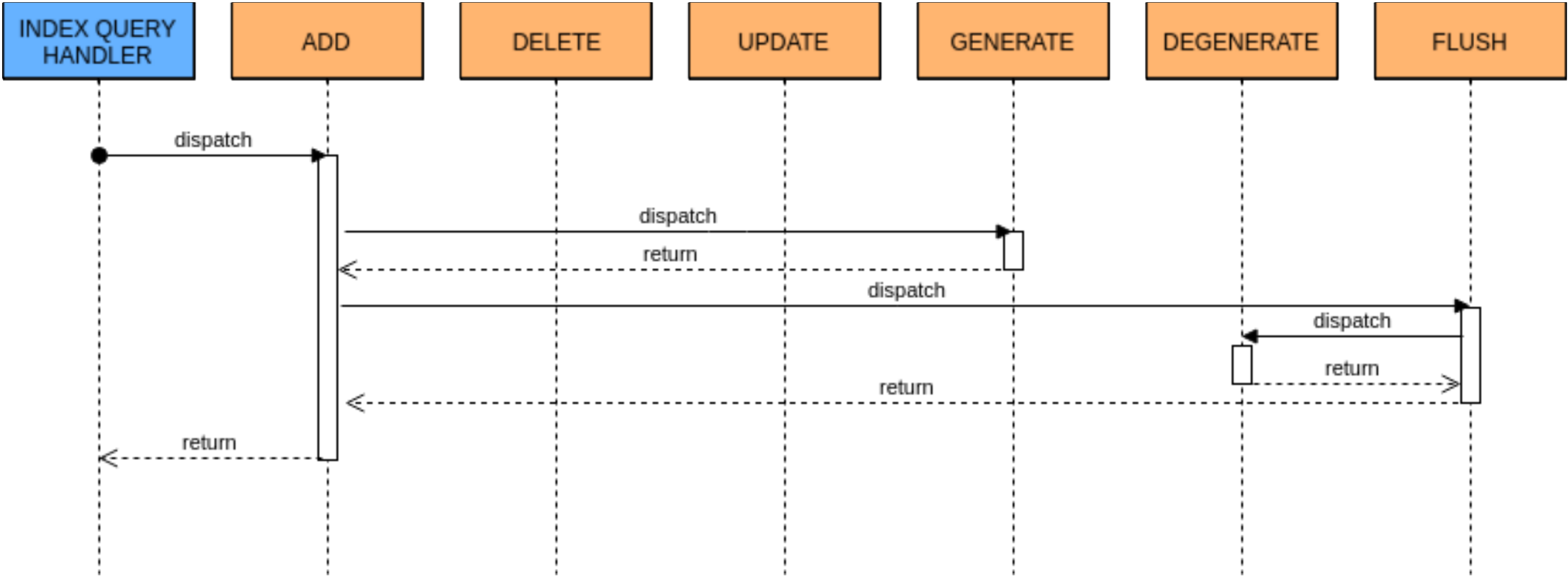}
\par Figure 2. Add Operation
\end{center}
\item \href{https://github.com/apsdehal/structured-query-engine/blob/master/app/indexer/Indexer.py\#L71}{Delete}:
\newline
Deletion of a document takes place in a lazy approach. First, the document is marked for deletion. So any further requests when the document is still in the system can know that it is invalid. The $\_id$ and $\_type$ are stored in a list of documents to be deleted. It is cleaned up when the inverted index and document store are rebuilt. If the $\_type$, $\_id$ are invalid or the document is already marked for deletion the operation is returned as a failure. Finally, a $flush$ to file is called before returning the result of the operations a success.
\begin{center}
\includegraphics[width=10cm,height=10cm,keepaspectratio]{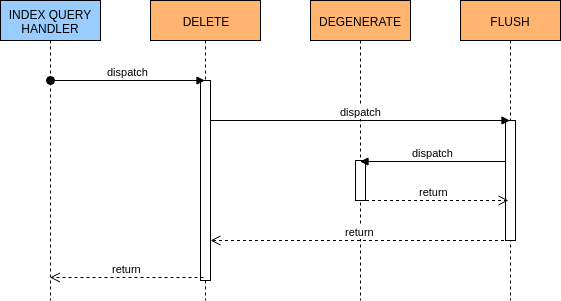}
\par Figure 3. Delete Operation
\end{center}
\item \href{https://github.com/apsdehal/structured-query-engine/blob/master/app/indexer/Indexer.py\#L47}{Update}:
\newline
Updating a document in SQE is very expensive. First, The old document if it exists is retrieved from the document store and is deleted immediately by by-passing the $delete$ function and calling the $degenerate$ function directly. Then the new document with the updated fields must be re-indexed in the same position by calling the $add$ function along with parameter for generating new $\_id$. The parameter is true if the old document does not exist.
\begin{center}
\includegraphics[width=13cm,height=10cm,keepaspectratio]{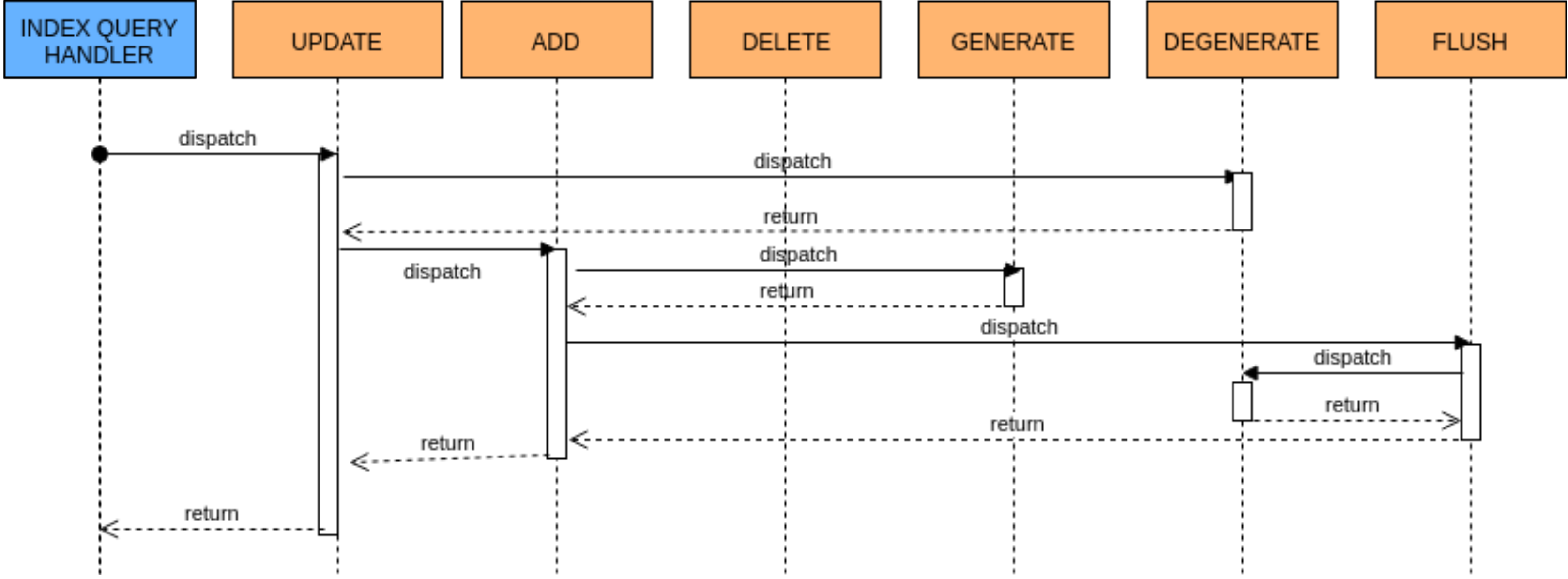}
\par Figure 4. Update Operation
\end{center}
\item \href{https://github.com/apsdehal/structured-query-engine/blob/master/app/indexer/Indexer.py\#L152}{Get}:
\newline
The get function checks if the $\_id$ is valid and it is not marked for deletion and then returns the corresponding document from the document store. Else, it returns an empty response.
\item \href{https://github.com/apsdehal/structured-query-engine/blob/master/app/indexer/Indexer.py\#L119}{Generate}:
\newline
The generate function is responsible to update both the inverted index and document stored.An inverted index is created for each $field$ under a $\_type$ in the mapping which has the $index$ field set to true. The term frequency is calculated using a $counter$ for each term in the $field$ and the shard number using the hash of the $\_id$. This is used to index the fields of the document. The document is added in the document store using its $\_id$ as the key.
\item \href{https://github.com/apsdehal/structured-query-engine/blob/master/app/indexer/Indexer.py\#L168}{Degenerate}:
\newline
The degenerate function takes the $\_id$ and $\_type$ pair and gets the corresponding document. The shard number is calculated using the hash of the $\_id$. This document is again tokenized after flattening it and instances of its fields are removed from the inverted index. The document is then removed from the document store using its $\_id$.
\item \href{https://github.com/apsdehal/structured-query-engine/blob/master/app/indexer/Indexer.py\#L208}{Flush}:
\newline
First, the $degenerate$ function is called to remove all the documents marked for deletion in the list of documents to be deleted and then the updated shards of inverted index and document store are stored. Flushing the contents of the memory to file ensures that all the components of the SQE share the same copy of the data. It is typically done after each $Add/Delete/Update$ command based on the $debounce$ time. 
\end{enumerate}
\subsubsection{Debounce}
Debouncing is the technique by which it is ensured that a function is executed only after a  certain amount of time from its previous call. Since the indexers and the retrievers share common data structures, it is important the they all access the same copy of the information. To ensure synchronization, each Add/Delete/Update command gives a call to flush the content to file. However, this call is debounced\footnote{\href{https://github.com/apsdehal/structured-query-engine/blob/master/app/helpers/utils/Debounce.py}{Code} for Debouncer.} to ensure that CPU utilization  and network bandwidth is optimized. 
\subsubsection{Compression}
All data structures are compressed before storage to minimize disk space and maximize network bandwidth usage. The compression\footnote{\href{https://github.com/apsdehal/structured-query-engine/blob/master/app/helpers/utils/Compressor.py}{Code} for Compressor.} is done using python-snappy which uses Google's snappy library internally for fast compression and decompression \cite{gunderson2015snappy}.

\subsection{Information Retriever}
\subsubsection{Structured Search}
Structured search\cite{TopDown} is about interrogating data that has inherent structure. By storing data in structured format we are not limited to search using a "simple string query". We can combine various string queries in complex structured query and can search data using this structured query \cite{Al-Khalifa:2003:QST:872757.872761}. A structured query is an Abstract Syntax Tree representation of a search expression, expressed in XML or JSON. Our structured Query Engine supports following queries.
\begin{enumerate}
\item Leaf Query (Term Query and Match query)
\newline
Leaf query is the most basic query type that can be used to compare a field (or fields) to a query string. The \textbf{Term Query} will search for the exact query that is specified, and does not analyze or tokenize the query. \textbf{Match query} will score all the docs against the query. Docs with highest score will be included in the results. Match query will be analyzed and tokenized before its searched in docs.

\item Compound Query (Bool Query)
\newline
Queries are seldom simple one-clause match queries. We frequently need to search for the same or different query strings in one or more fields, which means that we need to be able to combine multiple query clauses and their relevance scores in a way that makes sense. Bool query is composed of three sections:
\begin{enumerate}
\item should : At least one of these clauses must match. The equivalent of OR. 
\item filter : Clauses that must match, but are run in non-scoring, filtering mode.
\end{enumerate} 

\item Range Filter
\newline
Range filter can be applied to numerical fields and is often used to filter out results based on ranges. For example, only those movies which have imdb score greater than 8.0 are to be included in search results. Range filter can be passed as a parameter in compound query as shown in example below. Leaf query does not support range filter. The range query supports both inclusive and exclusive ranges, through combinations of the following options.
\begin{enumerate}
\item gt  : $>$ greater than
\item lt  : $<$ less than
\item gte : $\geq$ greater than or equal to
\item lte : $\leq$ less than or equal to

\end{enumerate}
\item Boosting Query Classes
\newline
While searching using compound query we can fine-tune the relevance score by providing more weight to a certain query clause. More weight means that documents containing that query will receive a higher relevance score than those that don't, which means that they will appear higher in the list of results. By default each query clause is assigned a weight of 1.0
\begin{center}
\begin{verbatim}
Example of compound query:
GET /_search
{"query": 
  {"bool": 
    {"should":
       [{"match": {"director_name": "Shane Black"}},
       {"match": {"plot_keywords": {"query": "human bomb", "boost": 2.0}}},
       {"match": {"actor_names": "Robert Downey Jr."}},
       {"term": {"actor_names": "Cheadle"}},
       {"match": {"genres": {"query": "Action", "boost": 2.0}}},
       {"match": {"genres": {"query": "Sci-Fi", "boost": 2.0}}}],
    "filter":
       [{"range": {"imdb_score": {"gte": 6.0}}}]
    }
  }
}
\end{verbatim}
\end{center}
\end{enumerate}
\begin{center}
\includegraphics[width=10cm,height=10cm,keepaspectratio]{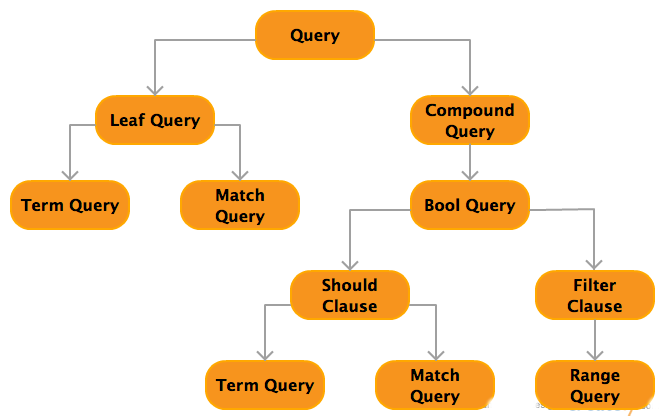}
\par Figure 2. Query DSL
\end{center}

\subsubsection{Scoring Documents} 
The multi-field search queries are combined query strings which are to be searched in one or more than one field \cite{Trotman2004619}. While scoring they are first decomposed back in multiple queries based on their search field, such that each query is to be searched in only one field. Then for each field present in query string we create a vector-space representation of the query using the document frequency table. Each dimension of the vector is set to the corresponding term's TF-IDF value. For a query clause without a boost value the TF value of the term is set to 1. For a query clause where boost value is provided we set the TF value equal to boost value.

Then we look up the postings list for each term in the query. Each document gets converted to its vector space representation. Again, each dimension is be set to the corresponding term's TF-IDF value. The documents are then scored. Each document's score is the inner product of its vector and the query vector. As the single multi-field query was decomposed into multiple queries, we repeat this step for all those queries and score from each step is added to get final score of each document. 

\section{Movie Recommendation System}

To demonstrate the working of SQE, we have built a movie recommendation system \footnote{Source code for movie recommendation system is available on \href{https://github.com/apsdehal/movie-recommendations}{Github} for reference} which recommends users movies based on a movie selected by them. In a typical movie recommendation application, user wants to be able to search for movie based on their preferences for director, actor, genres and movie score and then get similar movies. Search query provided by the user after putting his preferences becomes a complex query which can't be handled by simple term based search engine. Thus, a query engine is necessary which can search documents based on structure and particular pattern. In our movie recommendation application, built upon SQE, users are first able to search for a movie they like using various filters of director name, actor name, movie title, genres and IMDb score. Then, they are recommended similar movies on selecting a movie. These recommendations are also retrieved from the SQE itself. IMDB 5000 movie dataset on Kaggle was used to build this movie recommendation system \cite{IMDB5000}.

\subsection{Architecture} Movie recommendation system can be divided into three main components, namely (1) Frontend (2) Search Controller (3) Recommendation Controller. These are explained in detailed:

\begin{center}
\includegraphics[width=10cm,height=10cm,keepaspectratio]{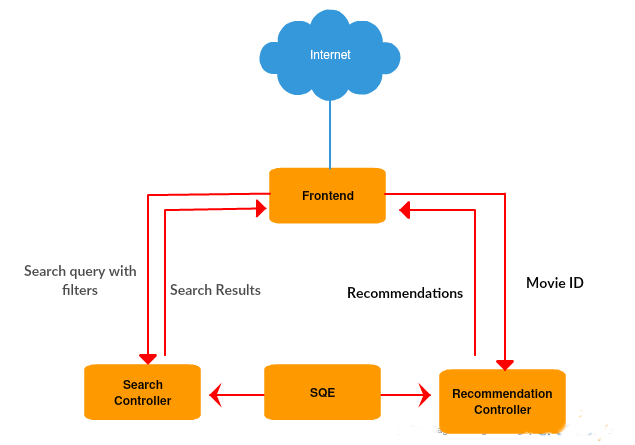}
\par Figure 3. Architecture of Movie Recommendation System
\end{center}
\subsection{Frontend} It is written in React.js \cite{Gackenheimer2015} and Bootstrap. It is a single page application with HTML 5 history support routing \cite{MozillaHistoryAPI} and asynchronous search and recommend. Frontend has two main routes (react components), home and movie. In home, a search form is presented with option to select from movie title, director name and actor name with a query box to fill the search query related to selected option. Genres can be selected in the next dropdown and finally minimum IMDb score for search result can be set. In movie section, search form is present with a selected movie and a panel with  recommended movies for that particular movie is present. Axios \footnote{More information on Axios on \href{https://github.com/mzabriskie/axios}{Github}} package is used for making asynchronous HTTP requests to backend.

\subsection{Backend}

Backend for the movie recommendation system is based on MVC (Model View Controller) \cite{deacon2009model} design pattern written in Tornado web framework. There are two controllers each for search and recommendations, which both have underlying models which interact with SQE through HTTP requests. Both of these controllers handle different routes, get the query arguments and pass the formatted parameters to respective models to return back the results as JSON in response.

\subsubsection{Search:} 
Search controller parses the complex information about the required movie from the user into a structured query as handled by SQE. Controller decodes the HTTP request and passes the parameters to the "query model" \footnote{See \href{https://github.com/apsdehal/movie-recommendations/blob/master/app/models/QueryModel.py\#L11}{code} for the query model for reference}. Query model builds up a structured query object like that of Section 2.4.2. One or more of "should" fields should match the results, whether on actor name, genres or IMDb score. If more than one of should filters match with a document, the score of the documents goes relatively high than other documents giving that document higher ranking in search results. IMDb score filter is implemented using "range" query using "gte" filter for imdb score field. Results are returned by SQE in the format as that of Elasticsearch, which are then displayed to user using Thumbnail \footnote{\label{footnote1} \href{https://github.com/apsdehal/movie-recommendations/blob/master/app/frontend/src/components/Thumbnail.js}{Source code} for Thumbnail component} component.

\subsubsection{Recommendation:}
Recommendation controller receives HTTP request containing document id of a movie from the frontend and it queries the SQE using this id to receive entire document containing movie information. Then it extracts the information of actors, directors, genres and plot keywords from the movie document. This information is passed as parameter to "Recommendation Model" \footnote{See \href{https://github.com/apsdehal/movie-recommendations/blob/master/app/models/RecommendationModel.py\#L7}{code} for the Recommendation model for reference}. Recommendation model builds up a structured query object from these parameters using compound query model. To fine-tune the results "plot keywords" and "genre" fields are given higher weights while building the query using boost clause. Results are returned by SQE in the format as that of Elasticsearch, which are then displayed to user using Thumbnail \textsuperscript{\ref{footnote1}} component.

\section{Conclusions}

As it can be seen through the example of movie recommendation engine, performing complex queries on data is possible through SQE. Fast retrieval, indexing, updation and deletion is made possible through smart inverted indices created for each field. Using techniques like compression through snappy and write (flush) limit through debouncing, we have made sure that memory space consumed is low and there are no excessive writes to the disk. Shared memory resources between retrievers and indexers promote efficient memory usage. With the help of boosting it was possible to create better recommendations for movies as movie's plot is more important for recommendations than movie's director or actor. NGram analyzer makes it possible to search on phrases from document tokens. Nested type are supported via flattening of mapping and at the same time number of shards are configurable which can be distributed to scale the whole system. Using Elasticsearch like query dsl it is possible to create queries for union and intersection of some particular sets. Further, everything is readily integrable into any project through RESTful API which can be queried using HTTP requests, thus SQE adds no library dependencies into projects that use it. \\

  	Even though, SQE implements a lot of features of Elasticsearch, there is still a room for lot of improvements. SQE in overall, presents a base for creating structured query engines like Elasticsearch. When the scale of documents will increase, SQE will have issues with memory bounds as currently it doesn't support proper memory allocation and garbage collections for documents loaded into memory. A better way to load documents would be to first remove documents from memory which are no longer needed according to a scheduling algorithm and then load more documents into memory. Other effective mechanisms are present to update the index efficiently \cite{Jang:1999:EMI:319950.320031}. As the index size considerably grows with number of documents, compression algorithms need to modified to adjust to scale. Our current compression system can be improved to use more efficient compression algorithms which rely on building inverted indices in different ways with tree (B-Trees) \cite{chen2010inverted} \cite{ziviani2000compression} \cite{moffat2000binary} \cite{melink2001building}. At the moment, query dsl of SQE only supports queries nested one level down which through recursive implementation can be used to create queries of any depth and any number of compound nodes. Elasticsearch aggregations\footnote{See elasticsearch \href{https://www.elastic.co/guide/en/elasticsearch/reference/current/search-aggregations.html}{documentation} for more information on aggregations} which can be used to fetch aggregate data based on a search query are also not supported by the SQE currently. Aggregation module can be built upon query dsl of SQE. Caching is not enabled at any place except for nginx in SQE. Scope of caching include caching filters which are mostly the same for queries of particular type and caching of simple queries which occur frequently. Another possibility is to use event based architecture to decouple indexers and retriever. \\

	SQE demonstrates how better search results can be provided on the basis of user preferences. Once we have a system to determine user's preferences, we can provide him/her personal search results. Simple term based matching doesn't help in long run with systems getting complex everyday and thus, need and role of SQE comes into play.


\begin{acks}

We would like to express our thanks to Professor Matt Doherty for his valuable and constructive suggestions during the planning and development of this project. His willingness to give his time so generously has been very much appreciated.

\end{acks}

\bibliographystyle{ACM-Reference-Format}
\bibliography{sample-bibliography}


\begin{thebibliography}{00}


\ifx \showCODEN    \undefined \def \showCODEN     #1{\unskip}     \fi
\ifx \showDOI      \undefined \def \showDOI       #1{#1}\fi
\ifx \showISBNx    \undefined \def \showISBNx     #1{\unskip}     \fi
\ifx \showISBNxiii \undefined \def \showISBNxiii  #1{\unskip}     \fi
\ifx \showISSN     \undefined \def \showISSN      #1{\unskip}     \fi
\ifx \showLCCN     \undefined \def \showLCCN      #1{\unskip}     \fi
\ifx \shownote     \undefined \def \shownote      #1{#1}          \fi
\ifx \showarticletitle \undefined \def \showarticletitle #1{#1}   \fi
\ifx \showURL      \undefined \def \showURL       {\relax}        \fi
\providecommand\bibfield[2]{#2}
\providecommand\bibinfo[2]{#2}
\providecommand\natexlab[1]{#1}
\providecommand\showeprint[2][]{arXiv:#2}

\bibitem[\protect\citeauthoryear{??}{W3R}{2004}]%
        {W3Rest}
 \bibinfo{year}{2004}\natexlab{}.
\newblock \bibinfo{title}{{"Web Services Architecture"}. World Wide Web
  Consortium. 11 February 2004. 3.1.3 Relationship to the World Wide Web and
  REST Architectures}.
\newblock
  \bibinfo{howpublished}{\url{https://www.w3.org/TR/2004/NOTE-ws-arch-20040211/\#relwwwrest}}.
    (\bibinfo{year}{2004}).
\newblock
\newblock
\shownote{Retrieved: 2016-04-30.}


\bibitem[\protect\citeauthoryear{??}{Bot}{2013}]%
        {BottomUp}
 \bibinfo{year}{2013}\natexlab{}.
\newblock \bibinfo{title}{Elasticsearch from the Bottom Up, Part 1}.
\newblock
  \bibinfo{howpublished}{\url{https://www.elastic.co/blog/found-elasticsearch-from-the-bottom-up}}.
    (\bibinfo{year}{2013}).
\newblock
\newblock
\shownote{Accessed: 2017-04-10.}


\bibitem[\protect\citeauthoryear{??}{Top}{2014}]%
        {TopDown}
 \bibinfo{year}{2014}\natexlab{}.
\newblock \bibinfo{title}{Elasticsearch from the Top Down}.
\newblock
  \bibinfo{howpublished}{\url{https://www.elastic.co/blog/found-elasticsearch-top-down}}.
    (\bibinfo{year}{2014}).
\newblock
\newblock
\shownote{Accessed: 2017-04-10.}


\bibitem[\protect\citeauthoryear{??}{IMD}{2016}]%
        {IMDB5000}
 \bibinfo{year}{2016}\natexlab{}.
\newblock \bibinfo{title}{IMDb 5000 Movies Dataset by Jerry Sun on Kaggle}.
\newblock
  \bibinfo{howpublished}{\url{https://www.kaggle.com/deepmatrix/imdb-5000-movie-dataset}}.
    (\bibinfo{year}{2016}).
\newblock
\newblock
\shownote{Accessed: 2017-04-10.}


\bibitem[\protect\citeauthoryear{??}{Moz}{2017}]%
        {MozillaHistoryAPI}
 \bibinfo{year}{2017}\natexlab{}.
\newblock \bibinfo{title}{Hisoty - Web APIs | MDN}.
\newblock
  \bibinfo{howpublished}{\url{https://developer.mozilla.org/en-US/docs/Web/API/History}}.
    (\bibinfo{year}{2017}).
\newblock
\newblock
\shownote{Retrieved: 2016-04-30.}


\bibitem[\protect\citeauthoryear{??}{Ela}{2017}]%
        {ElasticQueryDSL}
 \bibinfo{year}{2017}\natexlab{}.
\newblock \bibinfo{title}{{Query DSL} | Elasticsearch Reference [5.3]}.
\newblock
  \bibinfo{howpublished}{\url{https://www.elastic.co/guide/en/elasticsearch/reference/current/query-dsl.html}}.
    (\bibinfo{year}{2017}).
\newblock
\newblock
\shownote{Accessed: 2017-04-30.}


\bibitem[\protect\citeauthoryear{Al-Khalifa, Yu, and Jagadish}{Al-Khalifa
  et~al\mbox{.}}{2003}]%
        {Al-Khalifa:2003:QST:872757.872761}
\bibfield{author}{\bibinfo{person}{Shurug Al-Khalifa}, \bibinfo{person}{Cong
  Yu}, {and} \bibinfo{person}{H.~V. Jagadish}.}
  \bibinfo{year}{2003}\natexlab{}.
\newblock \showarticletitle{Querying Structured Text in an XML Database}. In
  \bibinfo{booktitle}{{\em Proceedings of the 2003 ACM SIGMOD International
  Conference on Management of Data}} {\em (\bibinfo{series}{SIGMOD '03})}.
  \bibinfo{publisher}{ACM}, \bibinfo{address}{New York, NY, USA},
  \bibinfo{pages}{4--15}.
\newblock
\showISBNx{1-58113-634-X}
\showDOI{%
\url{https://doi.org/10.1145/872757.872761}}


\bibitem[\protect\citeauthoryear{Brin and Page}{Brin and Page}{1998}]%
        {Brin:1998:ALH:297810.297827}
\bibfield{author}{\bibinfo{person}{Sergey Brin} {and} \bibinfo{person}{Lawrence
  Page}.} \bibinfo{year}{1998}\natexlab{}.
\newblock \showarticletitle{The Anatomy of a Large-scale Hypertextual Web
  Search Engine}.
\newblock \bibinfo{journal}{{\em Comput. Netw. ISDN Syst.\/}}
  \bibinfo{volume}{30}, \bibinfo{number}{1-7} (\bibinfo{date}{April}
  \bibinfo{year}{1998}), \bibinfo{pages}{107--117}.
\newblock
\showISSN{0169-7552}
\showDOI{%
\url{https://doi.org/10.1016/S0169-7552(98)00110-X}}


\bibitem[\protect\citeauthoryear{Chen, Tsai, Chandrasekhar, Takacs, Vedantham,
  Grzeszczuk, and Girod}{Chen et~al\mbox{.}}{}]%
        {chen2010inverted}
\bibfield{author}{\bibinfo{person}{David~M Chen}, \bibinfo{person}{Sam~S Tsai},
  \bibinfo{person}{Vijay Chandrasekhar}, \bibinfo{person}{Gabriel Takacs},
  \bibinfo{person}{Ramakrishna Vedantham}, \bibinfo{person}{Radek Grzeszczuk},
  {and} \bibinfo{person}{Bernd Girod}.}
\newblock \showarticletitle{Inverted Index Compression for Scalable Image
  Matching.}
\newblock


\bibitem[\protect\citeauthoryear{Deacon}{Deacon}{2009}]%
        {deacon2009model}
\bibfield{author}{\bibinfo{person}{John Deacon}.}
  \bibinfo{year}{2009}\natexlab{}.
\newblock \showarticletitle{Model-view-controller (mvc) architecture}.
\newblock \bibinfo{journal}{{\em Online][Citado em: 10 de mar{\c{c}}o de 2006.]
  http://www. jdl. co. uk/briefings/MVC. pdf\/}} (\bibinfo{year}{2009}).
\newblock


\bibitem[\protect\citeauthoryear{Gackenheimer}{Gackenheimer}{2015}]%
        {Gackenheimer2015}
\bibfield{author}{\bibinfo{person}{Cory Gackenheimer}.}
  \bibinfo{year}{2015}\natexlab{}.
\newblock \bibinfo{booktitle}{{\em What Is React?}}
\newblock \bibinfo{publisher}{Apress}, \bibinfo{address}{Berkeley, CA},
  \bibinfo{pages}{1--20}.
\newblock
\showISBNx{978-1-4842-1245-5}
\showDOI{%
\url{https://doi.org/10.1007/978-1-4842-1245-5_1}}


\bibitem[\protect\citeauthoryear{Gormley and Tong}{Gormley and Tong}{2015}]%
        {Gormley:2015:EDG:2904394}
\bibfield{author}{\bibinfo{person}{Clinton Gormley} {and}
  \bibinfo{person}{Zachary Tong}.} \bibinfo{year}{2015}\natexlab{}.
\newblock \bibinfo{booktitle}{{\em Elasticsearch: The Definitive Guide\/}
  (\bibinfo{edition}{1st} ed.)}.
\newblock \bibinfo{publisher}{O'Reilly Media, Inc.}
\newblock
\showISBNx{1449358543, 9781449358549}


\bibitem[\protect\citeauthoryear{Gunderson}{Gunderson}{2015}]%
        {gunderson2015snappy}
\bibfield{author}{\bibinfo{person}{SH Gunderson}.}
  \bibinfo{year}{2015}\natexlab{}.
\newblock \showarticletitle{Snappy: A fast compressor/decompressor}.
\newblock \bibinfo{journal}{{\em code. google. com/p/snappy\/}}
  (\bibinfo{year}{2015}).
\newblock


\bibitem[\protect\citeauthoryear{Jang, Kim, and Shin}{Jang
  et~al\mbox{.}}{1999}]%
        {Jang:1999:EMI:319950.320031}
\bibfield{author}{\bibinfo{person}{Hyunchi Jang}, \bibinfo{person}{Youngil
  Kim}, {and} \bibinfo{person}{Dongwook Shin}.}
  \bibinfo{year}{1999}\natexlab{}.
\newblock \showarticletitle{An Effective Mechanism for Index Update in
  Structured Documents}. In \bibinfo{booktitle}{{\em Proceedings of the Eighth
  International Conference on Information and Knowledge Management}} {\em
  (\bibinfo{series}{CIKM '99})}. \bibinfo{publisher}{ACM},
  \bibinfo{address}{New York, NY, USA}, \bibinfo{pages}{383--390}.
\newblock
\showISBNx{1-58113-146-1}
\showDOI{%
\url{https://doi.org/10.1145/319950.320031}}


\bibitem[\protect\citeauthoryear{Lee, Yoo, Yoon, and Berra}{Lee
  et~al\mbox{.}}{1996}]%
        {Lee:1996:ISS:226931.226950}
\bibfield{author}{\bibinfo{person}{Yong~Kyu Lee}, \bibinfo{person}{Seong-Joon
  Yoo}, \bibinfo{person}{Kyoungro Yoon}, {and} \bibinfo{person}{P.~Bruce
  Berra}.} \bibinfo{year}{1996}\natexlab{}.
\newblock \showarticletitle{Index Structures for Structured Documents}. In
  \bibinfo{booktitle}{{\em Proceedings of the First ACM International
  Conference on Digital Libraries}} {\em (\bibinfo{series}{DL '96})}.
  \bibinfo{publisher}{ACM}, \bibinfo{address}{New York, NY, USA},
  \bibinfo{pages}{91--99}.
\newblock
\showISBNx{0-89791-830-4}
\showDOI{%
\url{https://doi.org/10.1145/226931.226950}}


\bibitem[\protect\citeauthoryear{Melink, Raghavan, Yang, and
  Garcia-Molina}{Melink et~al\mbox{.}}{2001}]%
        {melink2001building}
\bibfield{author}{\bibinfo{person}{Sergey Melink}, \bibinfo{person}{Sriram
  Raghavan}, \bibinfo{person}{Beverly Yang}, {and} \bibinfo{person}{Hector
  Garcia-Molina}.} \bibinfo{year}{2001}\natexlab{}.
\newblock \showarticletitle{Building a distributed full-text index for the
  web}.
\newblock \bibinfo{journal}{{\em ACM Transactions on Information Systems
  (TOIS)\/}} \bibinfo{volume}{19}, \bibinfo{number}{3} (\bibinfo{year}{2001}),
  \bibinfo{pages}{217--241}.
\newblock


\bibitem[\protect\citeauthoryear{Moffat and Stuiver}{Moffat and
  Stuiver}{2000}]%
        {moffat2000binary}
\bibfield{author}{\bibinfo{person}{Alistair Moffat} {and} \bibinfo{person}{Lang
  Stuiver}.} \bibinfo{year}{2000}\natexlab{}.
\newblock \showarticletitle{Binary interpolative coding for effective index
  compression}.
\newblock \bibinfo{journal}{{\em Information Retrieval\/}} \bibinfo{volume}{3},
  \bibinfo{number}{1} (\bibinfo{year}{2000}), \bibinfo{pages}{25--47}.
\newblock


\bibitem[\protect\citeauthoryear{Trotman}{Trotman}{2004}]%
        {Trotman2004619}
\bibfield{author}{\bibinfo{person}{Andrew Trotman}.}
  \bibinfo{year}{2004}\natexlab{}.
\newblock \showarticletitle{Searching structured documents}.
\newblock \bibinfo{journal}{{\em Information Processing \& Management\/}}
  \bibinfo{volume}{40}, \bibinfo{number}{4} (\bibinfo{year}{2004}),
  \bibinfo{pages}{619 -- 632}.
\newblock
\showISSN{0306-4573}
\showDOI{%
\url{https://doi.org/10.1016/S0306-4573(03)00041-4}}


\bibitem[\protect\citeauthoryear{Ziviani, De~Moura, Navarro, and
  Baeza-Yates}{Ziviani et~al\mbox{.}}{2000}]%
        {ziviani2000compression}
\bibfield{author}{\bibinfo{person}{Nivio Ziviani}, \bibinfo{person}{E~Silva
  De~Moura}, \bibinfo{person}{Gonzalo Navarro}, {and} \bibinfo{person}{Ricardo
  Baeza-Yates}.} \bibinfo{year}{2000}\natexlab{}.
\newblock \showarticletitle{Compression: A key for next-generation text
  retrieval systems}.
\newblock \bibinfo{journal}{{\em Computer\/}} \bibinfo{volume}{33},
  \bibinfo{number}{11} (\bibinfo{year}{2000}), \bibinfo{pages}{37--44}.
\newblock


\bibitem[\protect\citeauthoryear{Zobel and Moffat}{Zobel and Moffat}{2006}]%
        {Zobel:2006:IFT:1132956.1132959}
\bibfield{author}{\bibinfo{person}{Justin Zobel} {and}
  \bibinfo{person}{Alistair Moffat}.} \bibinfo{year}{2006}\natexlab{}.
\newblock \showarticletitle{Inverted Files for Text Search Engines}.
\newblock \bibinfo{journal}{{\em ACM Comput. Surv.\/}} \bibinfo{volume}{38},
  \bibinfo{number}{2}, Article \bibinfo{articleno}{6} (\bibinfo{date}{July}
  \bibinfo{year}{2006}).
\newblock
\showISSN{0360-0300}
\showDOI{%
\url{https://doi.org/10.1145/1132956.1132959}}


\end{thebibliography}

\end{document}